# Spatial Resolution Enhancement of Remote Sensing Mine Images using Deep Learning Techniques

E. Zioga · A. Panagiotopoulou · M. Stefouli · E. Charou · L. Grammatikopoulos · E. Bratsolis · N. Madamopoulos

______________________________________________________________________

**Abstract**

Deep learning techniques are applied so as to increase the spatial resolution of Sentinel-2 satellite imagery, depicting the Amynteo lignite mine in Ptolemaida, Greece. Resolution enhancement by factors 2 and 4 as well as by factors 2 and 6 using Very-Deep Super-Resolution (VDSR) and DSen2 networks, respectively, provides fairly well results on Amynteo lignite mine images. Particularly, the aim of this research is to super-resolve (i) Sentinel-2 bands from 10m/pixel and 20m/pixel to 5m/pixel, that is even greater than the sensor's resolution, using VDSR network and (ii) Sentinel-2 lower-resolution bands from 20m/pixel and 60m/pixel to 10m/pixel using DSen2 network.

**Keywords:** Amynteo, mine surface, Sentinel-2 multispectral bands, spatial resolution enhancement, deep learning technique

______________________________________________________________________

## 1. Introduction

In our days, satellite imagery has been used in various applications, such as environmental monitoring, resource exploration, and disaster warning. Several widely used remote sensing satellites record multiple spectral bands with different spatial resolution mainly due to storage and transmission bandwidth restrictions, improved signal-to-noise ratio in some bands through larger pixels and specific bands that do not require high spatial resolution design (e.g. atmospheric corrections). MODIS, VIIRS, ASTER, Worldview-3 and Sentinel-2 (S2) are well-known multi-spectral, multi-resolution sensors with resolution differing by a factor of about 2-6 [1]. Super-Resolution (SR) technology can overcome the aforementioned limitations and provide finer spatial details than those captured by the original acquisition sensor, so as to obtain a complete data cube at the maximal and even greater sensor resolution [1-3]. At the present implementation, we focus on SR of S2 images.

S2 consists of two identical satellites, 2A and 2B, placed in the same sun-synchronous orbit, phased at 180° to each other, achieving thus frequent revisits. The sensor acquires 13 spectral bands in the visible, near infrared, and short wave infrared part of the spectrum with 10m, 20m and 60m resolution. S2 serves a wide range of applications related to Earth's land and coastal water and also provides information for agricultural and forestry practices and for helping manage food security. Satellite images are used to determine various plant indices, such as leaf area chlorophyll and water content indexes. In addition to monitoring plant growth, S2 can be used to map changes in land cover and to monitor the world's forests. It also provides information on pollution in lakes and coastal waters. Images of floods, volcanic eruptions and landslides contribute to disaster mapping and help humanitarian relief efforts [4].

Towards the spatial resolution enhancement, early methods include interpolation resulting in images with little additional information content. Recent works have focused on deep learning (DL) techniques in order to upsample the low-resolution (LR) images and output more detailed depicted areas [1-2]. These techniques consider the prediction of the high-resolution (HR) images as a supervised machine learning problem with the relation between lower-resolution input to higher-resolution output being not explicitly specified, but learned from the example data. DL-based approaches capture much more complex and general relations, but in turn require massive amounts of training data, and large computational resources to solve the underlying, extremely high-dimensional and complex optimization problem.

The study area of this work is the Amynteo lignite mine in Ptolemaida, Greece. The area of Amynteo is of particular interest mainly after the landslide that occurred on June 10, 2017, resulting in a significant malfunction and financial blow to the Public Power Plant. Our objective is to increase the spatial resolution of S2 mine images, so that more distinct details in the depicted area could be provided. In this way, mapping the area becomes easier and long-term management as well as monitoring of mining area are achieved [3].

## 2. Methodology

In the present study, the Very Deep Super-Resolution (VDSR) [1] and the DSen2 [2] networks are developed and trained on S2 Amynteo lignite mine images. These techniques are analysed below.

*2.1 VDSR technique*

VDSR is a convolutional neural network architecture designed to perform single image SR [1], Figure 1. The VDSR network learns the mapping between low- and high-resolution images. This mapping is possible because LR and HR images have similar image content and differ primarily in high-frequency details. VDSR employs a residual learning strategy, meaning that the network learns to estimate a residual image. In the context of SR, a residual image is the difference between a HR reference image and a LR image that has been upscaled using bicubic interpolation to match the size of the reference image. The residual image contains information about the high-frequency details of the image. The VDSR network detects the residual image from the luminance of a color image. The luminance channel of an image, Y, represents the brightness of each pixel through a linear combination of the red, green, and blue pixel values. VDSR is trained using only the luminance channel because human perception is more sensitive to changes in brightness than to changes in color. After the VDSR network learns to estimate the residual image, by adding the estimated residual image to the upsampled LR image through bicubic interpolation, HR images can be reconstructed and then converted back to the RGB color space.

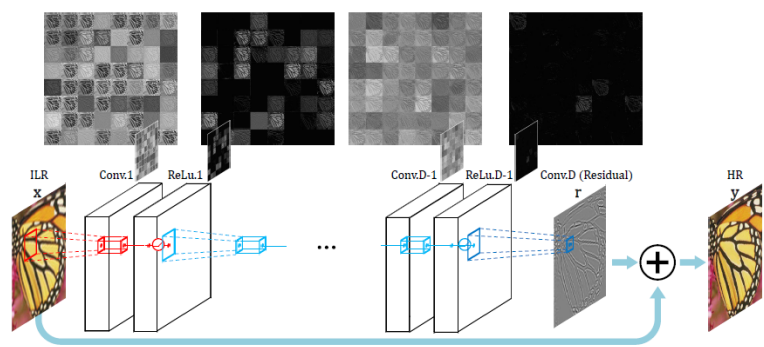

*Figure 1: The structure of the VDSR network [2].*

Particularly at the present work, VDSR technique is used to super-resolve S2 images of the spectral bands B2, B3, B4, B5, B6, B7, B8, B8A, B11, B12 to the increased spatial resolution of 5m/pixel.

*2.2 DSen2 technique*

DSen2 is also a convolutional neural network architecture based on two separate networks for different SR factors, since the 60m/pixel S2 bands cannot contribute information to the upsampling from 20 to 10m/pixel [2], Figure 2. The first network $T_{2x}$ upsamples the initial bands at 20m/pixel using information from bands at 10m/pixel and 20m/pixel that have been upscaled using bicubic interpolation to match the size of the reference bands. The second network $S_{6x}$ upsamples the initial bands at 60m/pixel using information from bands at 10m/pixel, 20m/pixel and 60m/pixel that have been upscaled using bicubic interpolation to match the size of the reference bands.

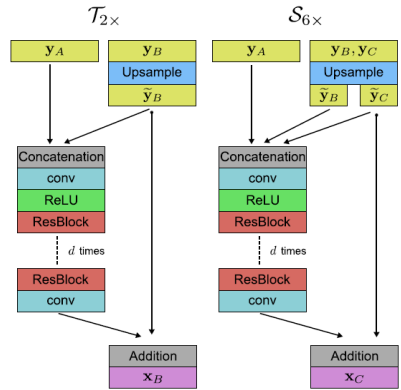

*Figure 2: The structure of the DSen2 network [1].*

In the context of this study, the DSen2 technique is applied to enhance the spatial resolution of S2 bands to 10m/pixel. In particular, the spatial resolution of B5, B6, B7, B8A, B11 and B12 bands (20m/pixel) is improved by a factor of 2, using information from B2, B3, B4 and B8 bands at 10m/pixel, and also information from the upscaled B5, B6, B7, B8A, B11 and B12 bands using bicubic interpolation from the initial resolution of 20m/pixel to 10m/pixel. In addition, the spatial resolution of B1, B9 bands (60m/pixel) is increased by a factor of 6, using information from B2, B3, B4 and B8 bands at 10m/pixel and also information from the upscaled B1, B5, B6, B7, B8A, B9, B11 and B12 bands using bicubic interpolation from the initial resolutions of 20m/pixel and 60m/pixel to 10m/pixel.

3. **Experimental Results**

Figure 3 shows the results of applying the VDSR technique to the S2 images of bands B2, B5, B8, B11 and B12. The spatial resolution of 10m/pixel and 20m/pixel is increased to 5m/pixel, so the original resolution is improved by factors 2 and 4, respectively. Carefully observing the images, more details of the mine surface are depicted at 5m/pixel images than in the original images of lower spatial resolution.

Figure 4 shows the results of applying the DSen2 technique to the S2 images of bands B1, B5, B9, B11 and B12. The resolution of all images is increased to 10m/pixel, thereafter the original resolution is improved by factors 2 and 6. In particular, observing the upscaled images of spectral bands B1 and B9, more details of the mine area are provided after the enhancement of the spatial resolution by a factor of 6.

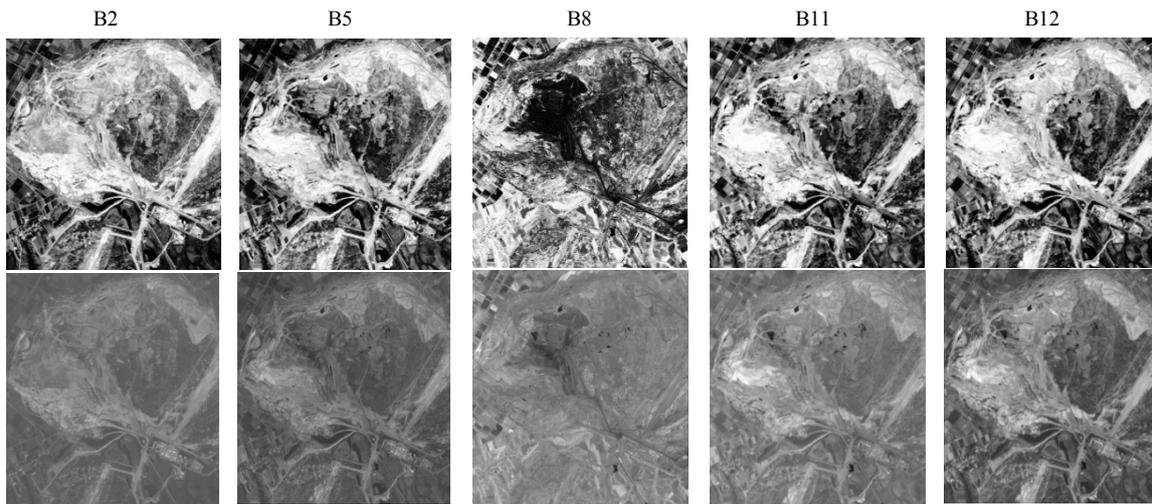

*Figure 3: The upscaled images at 5m/pixel using VDSR (above) and the original images at 10m/pixel (B2,B8) and 20m/pixel (B5,B11,B12) (below).*

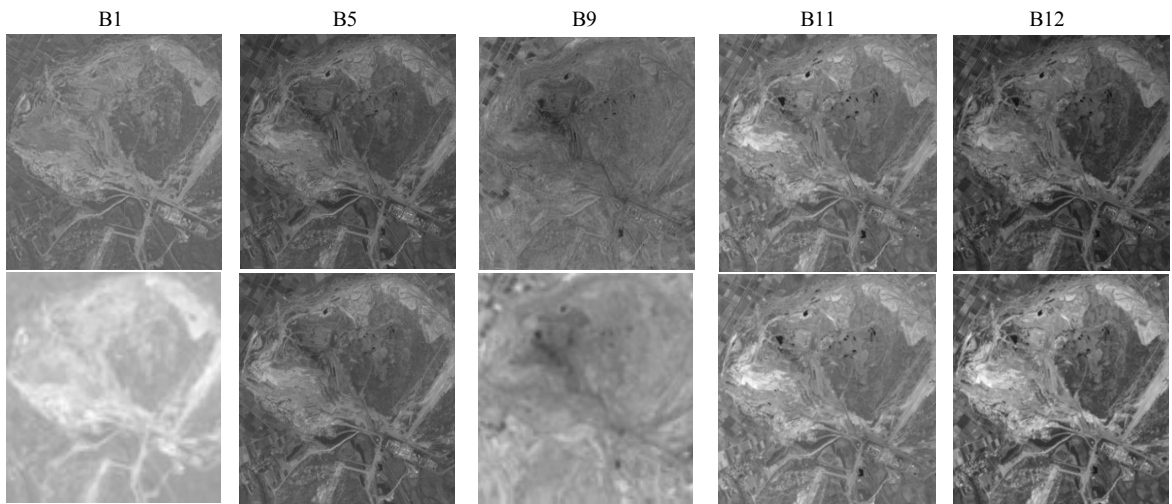

*Figure 4: The upscaled images at 10m/pixel using DSen2 (above) and the original images at 20m/pixel (B5,B11,B12) and 60m/pixel (B1,B9) (below).*

## 4. Conclusion

To conclude, VDSR and DSen2 networks have achieved very good results on Amynteo lignite mine images, with the upscaled images being more detailed and discrete than the original ones. Specifically, in the case of B1 and B9 bands, the difference between the output upscaled images and the LR ones at 60m/pixel is significant. In the future, we aim at applying additional single image DL SR techniques based on generative adversarial network (GAN) architecture as well as multi-image multi-spectral SR techniques, such as Brodu software. Eventually, the output upscaled images of the mine area shall be used in classification applications.